\documentclass[a4paper,showpacs,preprintnumbers,superscriptaddress,aps,prd,nofootinbib,floatfix,10pt]{revtex4-2}
\usepackage{graphicx}
\usepackage{dcolumn}
\usepackage{bm}
\usepackage{amsmath,amssymb,amsfonts}
\usepackage{latexsym}
\usepackage{color}
\usepackage{subcaption}
\usepackage{ulem}
\usepackage{mathrsfs}
\usepackage{braket}
\usepackage{yhmath}
\usepackage{cancel}
\usepackage{tensor}

\usepackage[usestackEOL]{stackengine}
 
\usepackage{multirow}

\usepackage{physics}
\usepackage{titlesec}

\renewcommand{\thesection}{\arabic{section}}
\renewcommand{\thesubsection}{\arabic{subsection}}

\usepackage{hyperref}
\hypersetup{%
setpagesize=false,
 bookmarksnumbered=true,%
 bookmarksopen=true,%
 colorlinks=true,%
 linkcolor=blue,
 citecolor=magenta,
}

\usepackage{graphicx}
\usepackage{dcolumn}
\usepackage{bm}
\usepackage{physics,amsmath,amssymb,amsfonts}
\usepackage{latexsym}
\usepackage{color}
\usepackage{slashed}

\def\nn    {\nonumber}
\DeclareUnicodeCharacter{039B}{\Lambda}

\usepackage{titlesec}

\titleformat{\section}
  {\normalfont\large\bfseries}{\thesection}{1em}{}
\titlespacing*{\section}
  {0pt}{1.ex plus .1ex minus .1ex}{0.ex plus .1ex minus .1ex}

\makeatletter
\renewcommand{\p@subsection}{\thesection.}
\renewcommand{\p@subsubsection}{\thesection.\thesubsection.}
\makeatother

\titleformat{\subsubsection}
  {\normalfont\itshape}{\thesection.\thesubsection.\thesubsubsection}{1em}{}
\titlespacing*{\subsubsection}
  {0pt}{1.ex plus .1ex minus .1ex}{0.ex plus .1ex minus .1ex}

\renewcommand{\thesubsection}{.\arabic{subsection}}

\begin{document}
\allowdisplaybreaks
\flushbottom
\title{\boldmath 
Nonminimal couplings and preheating effects in $R^2$-Higgs inflation after ACT and SPT data}

\author{Haneesh Gonuguntla}
\author{Tanmoy Modak}
\author{Arnab Samanta}

\affiliation{Department of Physical Sciences, Indian Institute of Science Education and Research Berhampur, Berhampur 760003, Odisha, India\\[0.1cm]}

\begin{abstract}
We study the effects of dimension-four and dimension-six nonminimal Higgs couplings to the Ricci scalar $R$ in the $R^2$-Higgs inflation model in light of the recent ACT and SPT observations. We show that the dimension-six operators $|\Phi|^2 R^2$ and $|\Phi|^4 R$ can accommodate the enhanced scalar spectral index $n_s$ preferred by the combined CMB+BAO analyses. Using a doubly covariant formalism, we find that the same region of parameter space that explains the observed value of $n_s$ can also induce rapid preheating through the production of the Goldstone modes. If thermalization proceeds efficiently through this preheating mechanism, it may help match the inflationary scale with the CMB reference scale.
\end{abstract}

\maketitle
\vspace{2ex}

\hrule
\vspace{2ex}
 \tableofcontents
\vspace{2ex}
\numberwithin{equation}{section}
\hrule
\setlength{\parskip}{1\baselineskip}
\setlength{\parindent}{0pt}
\vspace{1ex}
\section{Introduction}\label{sec:intro}
Recent measurements of the scalar spectral index $n_s$ from combined cosmic microwave background (CMB) observations by Planck~\cite{Planck:2018jri}, the Atacama Cosmology Telescope (ACT)~\cite{AtacamaCosmologyTelescope:2025blo}, and the South Pole Telescope (SPT)~\cite{SPT-3G:2025bzu}, together with baryon acoustic oscillation (BAO) data from DESI, disfavor the original Starobinsky ($R^2$) inflation model~\cite{Starobinsky:1980te,Starobinsky:1983zz,Vilenkin:1985md,Mijic:1986iv,Maeda:1987xf} at the $\sim 2\sigma$ level~\cite{SPT-3G:2025bzu,AtacamaCosmologyTelescope:2025blo}. The same is true for the single-field like regime of the $R^2$-Higgs inflation model. Within the baseline $\Lambda$CDM model, the CMB-only determinations of $n_s$ remain mutually consistent: Planck finds $n_s = 0.9657 \pm 0.0040$, SPT-3G D1 finds $n_s = 0.951 \pm 0.011$, ACT DR6 finds $n_s = 0.9682 \pm 0.0069$, while the combined analyses yield $n_s = 0.9671 \pm 0.0058$ for ACT+SPT and $n_s = 0.9636 \pm 0.0035$ for Planck+SPT. However, the inclusion of BAO measurements from DESI DR2~\cite{DESI:2025zgx} shifts the preferred value of $n_s$ upward. In particular, ACT+DESI yields $n_s = 0.9752 \pm 0.0030$ (denoted P-ACT-LB2), while the combined CMB-SPA+DESI analysis gives $n_s = 0.9728 \pm 0.0027$.

For single-field attractor-type models, the scalar spectral index is approximately given by $n_s = 1 - 2/\mathcal{N}_*$, where $\mathcal{N}_*$ denotes the number of $e$-folding between the horizon exit of the reference scale, $k_{\rm ref}=0.05\,{\rm Mpc}^{-1}$, and the end of inflation. The value of $\mathcal{N}_*$ depends on the post-inflationary reheating history and plays a crucial role in matching inflationary scales to CMB observables. In $R^2$ inflation and in the single-field like regime of the $R^2$-Higgs model, $\mathcal{N}_*$ typically lies in the range $50$--$60$, yielding $n_s \simeq 0.9600$--$0.9667$. This discrepancy between the measured $n_s$ by ACT and SPT with BAO data denoted as the so called CMB-BAO tension~\cite{Drees:2025ngb,Kallosh:2025ijd}. Accommodating a larger value of $n_s$ in general requires $\mathcal{N}_*>60$, which may in turn necessitate nontrivial reheating mechanisms that are not typically realized in either $R^2$ or the $R^2$-Higgs model. Although one may invoke an exotic reheating mechanism within these models, several alternatives have recently been proposed~\cite{Drees:2025ngb,Kallosh:2025ijd,Aoki:2025wld,Gialamas:2025kef,Zharov:2025evb,Liu:2025qca,Haque:2025uis,Yogesh:2025wak,Gialamas:2025ofz,Addazi:2025qra,Pallis:2025nrv,Saini:2025jlc,Wolf:2025ecy,Wang:2025dbj,Piva:2025cqi,SidikRisdianto:2025qvk,Zharov:2025zjg,Ferreira:2025lrd,Ellis:2025ieh,Odintsov:2025jky,Oikonomou:2025htz,Park:2025upd,Odintsov:2025eiv,Pozdeeva:2025wsl,Modak:2025bjv,Modak:2025grj,Pineda:2025ubm,Nojiri:2026hij,Odintsov:2026doe,Oikonomou:2026mvp,Odintsov:2026cxz,Odintsov:2026dss} to account the tension. These scenarios can accommodate larger values of $n_s$ while keeping $\mathcal{N}_*$ within the conventional range of $\left[50,60\right]$ $e$-folding.

In this article, we focus on the single-field like regime of the $R^2$-Higgs model and show that the dimension-six operators $(1/\xi_1)|\Phi|^2 R^2$ and $(1/\xi_2)|\Phi|^4 R$, where $\Phi$ denotes the Higgs doublet, can each accommodate the high value of $n_s$ found by the latest CMB+BAO data while keeping $\mathcal{N}_*$ within $50$--$60$ $e$-folding. Such higher dimensional operators naturally arise in general $f(\Phi,R)$ theories and provide a powerful probe of the UV sensitivity of the model~\cite{Jinno:2019und}. They can modify inflationary observables, including $n_s$, the tensor-to-scalar ratio $r$, and the scalar amplitude $\mathcal{A}_s$~\cite{Lee:2023wdm,Jinno:2019und}. We show that the nonminimal couplings $\xi_1$ and $\xi_2$, associated with the dimension-six operators, can account for the observed value of $n_s$ for $\xi_H \gtrsim 1$, where $\xi_H$ denotes the dimension-four nonminimal coupling associated with the operator $\xi_H |\Phi|^2 R$.

We also find that, while the parameters $\xi_1$ and $\xi_2$ may help account for the inflationary observables, $\xi_H$ may help match the CMB and inflationary scales through preheating. We show that $\xi_H \gtrsim 1$ can induce successful preheating from the Goldstone boson sector, which helps match the CMB scale to the inflationary scale. This effect has been largely missing from the literature in the context of $R^2$-Higgs inflation due to the commonly used unitary gauge choice. We discuss how a proper gauge choice, such as the Coulomb gauge, in which the Goldstone bosons (or equivalently the longitudinal gauge boson) are dynamical, can lead to rapid preheating after inflation for $\xi_H \gtrsim 1$. If thermalization occurs immediately after preheating, this mechanism may help bring $\mathcal{N}_*$ below 60 $e$-folding.

The primary aim of this work is to show that two qualitatively different classes of the Higgs-curvature operators $|\Phi|^2R^2$ and $|\Phi|^4R$, can both accommodate the observed inflationary observables. While the resulting inflationary predictions are practically identical to the case of $R^3$ modifications, the underlying dynamics are different. In the case of the $R^3$ correction, the enhancement of $n_s$ is independent of the Higgs background $h_0$ and can be realized even in the absence of the Higgs field. By contrast, for the $|\Phi|^2R^2$ and $|\Phi|^4R$ operators, the Higgs field plays an essential role in both enhancing $n_s$ and facilitating scale matching and, thereby emphasizing the role of Higgs field in accounting the ACT and SPT data. We further briefly discuss the possibility of distinguishing the $|\Phi|^2R^2$ and $|\Phi|^4R$ scenarios and the $R^3$ modifications in the context of $R^2$-Higgs inflation.

The paper is organized as follows. In Sec.~\ref{actio}, we present the action and derive the relevant equations of motion. The inflationary dynamics are discussed in Sec.~\ref{sec:infdynamics}, while preheating is analyzed in Sec.~\ref{sec:prehea}. We conclude with a discussion of our results in Sec.~\ref{disc}.

\section{Action and Equations of Motion}\label{actio}

The Jordan frame action of the $R^2$-Higgs model, including the dimension-six nonminimal couplings, is given by
\begin{align}
  S_J  =  \int d^4 x \sqrt{-g_J} \bigg[ \frac{M_{\rm P}^{2}}{2} f(R_J, \Phi)
  -g_J^{\mu\nu}(\nabla_\mu\Phi)^\dagger \nabla_\nu\Phi - V(\Phi, \Phi^\dagger)  -  \dfrac{1}{4} g_J^{\mu\rho} g_J^{\nu\sigma} B_{\mu\nu}B_{\rho\sigma}
  - \dfrac{1}{4} g_J^{\mu\rho} g_J^{\nu\sigma} W^i_{\mu\nu}W^i_{\rho\sigma}
 \bigg],\label{eq:action1}
\end{align}
where,
\begin{align}
    f(R_J,\Phi)&=R_J+\frac{\xi_R}{2M^2_{\rm P}}R^2_J+\frac{2\xi_H}{M^2_{\rm P}}\left|{\Phi}\right|^2R_J+\frac{1}{M_{\rm P}^4\xi_1} \left|{\Phi}\right|^2R_J^2 +\frac{1}{M_{\rm P}^4\xi_2} \left|{\Phi}\right|^4R_J, \ \ V(\Phi,\Phi^\dagger)=\lambda \left|{\Phi}\right|^4,
\label{eq:actionJ}
\end{align}
with $G$ being Newton's constant and $M_{\rm P}=\sqrt{1/\left(8\pi G\right)}\approx 2.4\times 10^{18}~\text{GeV}$. Throughout this work, we follow the mostly-plus convention where $\sqrt{-g_J}$ denoted as determinant of the metric $g_J^{\mu\nu}$ and the convention of Levi-Civita tensor is chosen as $\epsilon^{0123}=1$. Here $\Phi$ is Higgs doublet, $R_J$ is the space-time Ricci scalar in Jordan frame and, $B_{\mu\nu}$ and $W^i_{\mu\nu}$ are the field-stress tensors of the SM gauge groups respectively.
The covariant derivative for the $SU(2)_L \times U(1)_Y$ groups are defined as
\begin{align}
\nabla_\mu = D_\mu + i g' \frac{1}{2} Q_{Y_f} B_\mu + i g \, \bm{T} \cdot \bm{W}_\mu,
\end{align}
where $g'$ and $g$ respective gauge couplings, $\bm{T}={\bm{\tau}}/{2}$ are the weak-isospin and $Q_{Y_f}$ is the $U(1)_Y$ hypercharge with,  $\bm{\tau}$ are Pauli matrices. Note that the field-stress tensors of gauge fields are
\begin{align}
B_{\mu\nu} = D_\mu B_\nu - D_\nu B_\mu, \hspace{1.8cm} W^i_{\mu\nu} = D_\mu W^i_\nu - D_\nu W^i_\mu - g \sum_{j,k=1}^{3} \epsilon_{ijk} W^j_\mu W^k_\nu,
\end{align}
where $D_\mu$ is the covariant derivative appears space-time metric. While writing down the action in Eq.~\eqref{eq:actionJ} we have ignored the fermions but shall return to it shortly.

It is however convenient to study the inflationary and subsequent preheating dynamics of the $R^2$-Higgs model in the so called Einstein frame. We perform a frame transformation for a general $f(R_J,\Phi)$ theory by first introducing an auxiliary field $\Psi$ and then make a Legendre transformation. The action then becomes
\begin{equation} \begin{aligned}
  S_J  =&  \int d^4 x \sqrt{-g_J}  \bigg[\frac{M_{P}^{2}}{2} \left(f(\Psi,  \Phi)
  + \frac{\partial f(\Psi,  \Phi)}{\partial \Psi} (R_J-\Psi)\right)
   -g_J^{\mu\nu}(\nabla_\mu\Phi)^\dagger \nabla_\nu\Phi - V(\Phi, \Phi^\dagger) \nn\\
   &-  \dfrac{1}{4} g_J^{\mu\rho} g_J^{\nu\sigma} B_{\mu\nu}B_{\rho\sigma}
   - \dfrac{1}{4} g_J^{\mu\rho} g_J^{\nu\sigma} W^i_{\mu\nu}W^i_{\rho\sigma}\bigg]\label{eq:actionJ2}.
\end{aligned} \end{equation}
It is well-known that the Legendre transformation is well defined if $f(R,\Phi)$ is convex~\cite{Ivanov:2021chn}.
One may now introduce a physical degree of freedom
\begin{align}
\Theta = \frac{\partial f(\Psi,  \Phi)}{\partial \Psi},\label{eq:theta}
\end{align}
and with this newly introduced $\Theta$ the Eq.~\eqref{eq:actionJ2} becomes
\begin{equation} \begin{aligned}
  S_J  =  \int d^4 x \sqrt{-g_J} \bigg[& \frac{M_{P}^{2}}{2} \Theta R_J - U(\Theta,   \Phi)
  -g_J^{\mu\nu}(\nabla_\mu\Phi)^\dagger \nabla_\nu\Phi - V(\Phi, \Phi^\dagger)  \bigg].\label{eq:actionJ4}
\end{aligned} \end{equation}
The potential $U(\Theta,\Phi)$ is
\begin{align}
U(\Theta,   \Phi) &= \frac{M_{\rm P}^2}{2}\left[\Psi \Theta -  f(\Psi,  \Phi)\right] = \frac{M_{\rm P}^4 }{4 \xi_R }
\frac{\bigg(\Theta -1-2 \xi_H \frac{|\Phi|^2}{M_{\rm P}^2}-\frac{|\Phi|^4}{M_{\rm P}^4 \xi_2}\bigg)^2}{\bigg(1+  \frac{2|\Phi|^2}{M_{\rm P}^2 \xi_1 \xi_R}\bigg)}.
\end{align} In the limit $\xi_1\to \infty$ and $\xi_2 \to \infty$ the $U(\Theta,   \Phi)$ reduces to the one in $R^2$-Higgs one.
A Weyl rescaling of the metric $g^{\mu\nu}_J = \Theta \ g^{\mu\nu}_E$ leads to the action in the Einstein frame
\begin{equation} \begin{aligned}
S_E  = \int d^4 x \sqrt{-g_E}\bigg[ & \frac{M_{\rm P}^2}{2} R_E - \frac{3 M_{\rm P}^2}{4} g_E^{\mu \nu} \partial_\mu (\ln\Theta) \partial_\nu(\ln\Theta)
-\frac{1}{\Theta} g_E^{\mu\nu}(\nabla_\mu\Phi)^\dagger \nabla_\nu\Phi - V_E \\
&  -  \frac{1}{4} g_E^{\mu\rho} g_J^{\nu\sigma} B_{\mu\nu}B_{\rho\sigma}
  - \frac{1}{4} g_E^{\mu\rho} g_E^{\nu\sigma} W^i_{\mu\nu}W^i_{\rho\sigma}\bigg]\label{eq:actionE1},
\end{aligned} \end{equation}
where
\begin{align}
V_E &= \frac{1}{\Theta^2}\left[V(\Phi, \Phi^\dagger) +U(\Theta, \Phi)\right],\\
R_J &= \Theta \left[R_E +3 \Box_E  \Theta- \frac{3}{2} g_{E}^{\mu\nu} \partial_\mu (\ln\Theta) \partial_\nu(\ln\Theta) \right].
\end{align}
Here we have ignored the surface term in the action and $\Box_E = g_{E}^{\mu\nu} \partial_\mu \partial_\nu$.

A field redefinition and decomposition of Higgs field (with $Q_Y=+1$)
\begin{align}
\phi = M_{\rm P} \sqrt{\frac{3}{2}} \ln\Theta,~~~
\Phi =
\frac{1}{\sqrt{2}} \begin{pmatrix}
  \phi_3+ i \phi_4 \\
  h + i \phi_2 \\
\end{pmatrix}
\end{align}
leads to quadratic order
\begin{equation} \begin{aligned}
S_E  &= \int d^4 x \sqrt{-g_E}\bigg[\frac{M_{\rm P}^2}{2} R_E - \frac{1}{2} G_{IJ} g_E^{\mu \nu} D_\mu \phi^I D_\nu \phi^J - 
V_E(\phi^I) -  \dfrac{1}{4} g_E^{\mu\rho} g_E^{\nu\sigma} F_{A\mu\nu}F_{A\rho\sigma}-  
\dfrac{1}{4} g_E^{\mu\rho} g_E^{\nu\sigma} F_{Z\mu\nu}F_{Z\rho\sigma} \\
&  - \dfrac{1}{2} g_E^{\mu\rho} g_E^{\nu\sigma} F^+_{W\mu\nu}F^-_{W\rho\sigma} - e^{-\sqrt{\frac{2}{3}}\frac{\phi}{M_{\rm P}}} g_E^{\mu\nu}\bigg(\frac{g_Z^2}{8}h^2 Z_\mu Z_\nu+
\frac{g_Z}{2}\left[ (D_\mu h) \ \phi_2- (D_\mu\phi_2) h\right] Z_\nu +\frac{e^2}{4 s_W^2} h^2  W^+_\mu W^-_\nu+\\
&\frac{i e}{2 \sqrt{2} s_W} D_\mu h \left[W^-_\nu  (\phi_3+i \phi_4)  - W^+_\nu   (\phi_3-i \phi_4) \right]-
\frac{i e}{2 \sqrt{2} s_W}  \left[ W^-_\nu D_\mu(\phi_3+i\phi_4)  -  W^+_\nu D_\mu(\phi_3-i\phi_4) \right] h\bigg)\label{eq:actionfinal},
\end{aligned} \end{equation}
where $\phi^I \in \{\phi, h, \phi_2, \phi_3, \phi_4 \}$ and $g_Z= {e}/({s_W c_W})$. The  $G_{IJ}$s are the $5\times 5$ field-space metric   with non-vanishing elements $G_{\phi\phi} = 1$, $G_{ h h} = e^{-\sqrt{\frac{2}{3}}\frac{\phi}{M_{\rm P}}}$,$G_{\phi_i \phi_i} = e^{-\sqrt{\frac{2}{3}}\frac{\phi}{M_{\rm P}}}$ with $i=2,3,4$. The potential
\begin{align}
V_E = e^{-2\sqrt{\frac{2}{3}}\frac{\phi}{M_{\rm P}}}\left[\lambda \left|{\Phi}\right|^4 + \frac{M_{\rm P}^4 }{4 \xi_R }
\frac{\bigg(e^{\sqrt{\frac{2}{3}}\frac{\phi}{M_{\rm P}}} -1-2 \xi_H \frac{|\Phi|^2}{M_{\rm P}^2}-\frac{|\Phi|^4}{M_{\rm P}^4 \xi_2}\bigg)^2}{\bigg(1+  \frac{2|\Phi|^2}{M_{\rm P}^2 \xi_1 \xi_R}\bigg)}\right]\label{eq:potenein}
\end{align}
reduces to $R^2$-Higgs inflation in the limit $\xi_1,\xi_2\to \infty$
\begin{align}
V_E(\phi^I) =& e^{-2\sqrt{\frac{2}{3}}\frac{\phi}{M_{\rm P}}} \bigg[\frac{\lambda}{4} \left(h^2+\sum^4_{i=2} \phi_i^2\right)^2
+ \frac{M_{\rm P}^4}{4 \xi_R}\Biggl\{1 - e^{\sqrt{\frac{2}{3}}\frac{\phi}{M_{\rm P}}} + \frac{\xi_H}{M_{\rm P}^2} \left(h^2+\sum^4_{i=2} \phi_i^2\right)\Biggr\}^2\bigg]
\label{def:VE-final}.
\end{align}

Varying the quadratic action we get the equations of motion (EoMs) for the scalars fields $\phi^I$ as
\begin{equation} \begin{aligned}
&\Box \phi^K + \Gamma^{K}_{\ IJ} \ g_E^{\alpha \nu} D_\alpha  \phi^I D_\nu \phi^J - G^{KM} V_{E,M} \\&+ G^{KM}
e^{-\sqrt{\frac{2}{3}}\frac{\phi}{M_{\rm P}}} \left(\sqrt{\frac{2}{3}}\frac{1}{M_{\rm P}}\right) g_E^{\alpha\nu} (\partial_\alpha \phi)
\bigg(\frac{g_Z}{2} \delta^3_M h Z_\nu +\frac{i e}{2 \sqrt{2} s_W}  \left[ W^-_\nu  (\delta^4_M+i\delta^5_M)  -  W^+_\nu  (\delta^4_M-i\delta^5_M) \right] h\bigg)\\
&-G^{KM}e^{-\sqrt{\frac{2}{3}}\frac{\phi}{M_{\rm P}}} g_E^{\alpha\nu}
\bigg(\frac{g_Z}{2}   \ D_\alpha\left(\delta^3_M h Z_\nu\right)
+\frac{i e}{2 \sqrt{2} s_W}  D_\alpha\left[\left( W^-_\nu  (\delta^4_M+i\delta^5_M)  -  W^+_\nu  (\delta^4_M-i\delta^5_M) \right] h\right)\bigg)  \\
&-G^{KM}\bigg[e^{-\sqrt{\frac{2}{3}}\frac{\phi}{M_{\rm P}}} g_E^{\mu\nu}\delta^3_M\bigg(\frac{g_Z}{2}(D_\mu h) Z_\nu\bigg)+e^{-\sqrt{\frac{2}{3}}\frac{\phi}{M_{\rm P}}} g_E^{\mu\nu}\delta^4_M\bigg(\frac{i e}{2 \sqrt{2} s_W} D_\mu h \left(\ W^-_\nu - W^+_\nu \right)\bigg)\\
&+e^{-\sqrt{\frac{2}{3}}\frac{\phi}{M_{\rm P}}} g_E^{\mu\nu}\delta^5_M\bigg(\frac{i e}{2 \sqrt{2} s_W} D_\mu h \left(\ i W^-_\nu + i W^+_\nu \right)\bigg)\bigg]
= 0\label{eom:scalar},
\end{aligned} \end{equation}
where $\Gamma^{K}_{\ IJ}$ are the field-space Christoffel symbols. Note here that, we have only considered action at quadratic order for all fields in Eq.~\eqref{eq:actionfinal} since we restrict ourselves to linear approximation.

\section{Inflationary dynamics}
\label{sec:infdynamics}

\subsection{Background and perturbation}
We proceed towards the inflationary dynamics, we follow the covariant formalism as in Refs.~\cite{Gong:2011uw,Sfakianakis:2018lzf,Kaiser:2012ak} to capture multifield nature of the model with noncanonical kinetic terms. The fields $\phi^I(x^\mu)$ are decomposed into backgrounds (${\varphi}^I $) and perturbations ($\delta\phi^I$) as
\begin{align}
\phi^I(x^\mu) = \varphi^I(t) + \delta\phi^I(x^\mu)\label{fieldexpan},
\end{align}
with $t$ being the cosmic time. Here $\varphi^I(t) = \{\varphi(t),h_0(t)\}$, which are background fields for $\phi$ and $h$
respectively, while all other scalar fields including gauge bosons are treated as perturbations. The perturbed Friedmann-Robertson-Walker (FRW) metric in linear order is~\cite{Kodama:1984ziu,Mukhanov:1990me,Malik:2008im}
\begin{align}
ds^2 &= -(1+2\mathcal{A}) dt^2 +2 a(t) (\partial_i \mathcal{B}) dx^i dt +
a(t)^2 \left[(1-2\psi) \delta_{ij}+ 2 \partial_i \partial_j \mathcal{E}\right] dx^i dx^j,\label{eq:frwmetric}
\end{align}
with $\mathcal{A}, \mathcal{B}, \psi$ and $\mathcal{E}$ are the scalar metric perturbations and $a(t)$
is the scale factor. We choose the longitudinal gauge i.e. $\mathcal{E}=0$ and $ \mathcal{B}=0$ throughout our work.

Inserting the Eq.~\eqref{fieldexpan} in Eq.~\eqref{eom:scalar} we get EoMs for the background fields as
\begin{align}
&\mathcal{D}_t \dot{\varphi}^I + 3 H\dot{\varphi}^I + G^{\phi J} V_{E,J}(\varphi^I)= 0\label{eq:bkg_inf},
\end{align}
where field space covariant derivatives $\mathcal{D}_t$ and $\mathcal{D}_J$ are defined as~\cite{Gong:2011uw,Sfakianakis:2018lzf,Kaiser:2012ak}
\begin{align}
\mathcal{D}_t A^I &  = \dot{A}^I  + \Gamma^I_{\; JK} \dot{\varphi}^J A^K,\\
 \mathcal{D}_J A^I & = \partial_J A^I + \Gamma^I_{\; JK} A^K.
\end{align}
The Hubble parameter $H=\dot{a}/a$ can be expressed also as
\begin{align}
H^2 &= \left(\frac{\dot{a}}{a}\right)^2 = \frac{1}{3 M_{\rm P}^2} \bigg(\frac{1}{2} G_{IJ} \dot{\varphi}^I \dot{\varphi}^J + V_E(\varphi^I)\bigg).\label{hubble1}
\end{align}
The EoMs in Eq.~\eqref{eq:bkg_inf} are solved simultaneously with Eq.~\eqref{hubble1}. The $\dot{H}$ derived from Eq.~\eqref{hubble1} is compared with
$\dot{H} = -\frac{1}{2 M_{\rm P}^2} \bigg(G_{IJ} \dot{\varphi}^I \dot{\varphi}^J\bigg)$ for consistency. We denote the inflation ended  when $\epsilon = -\frac{\dot{H}}{H^2}$ equals to 1, which
corresponds to cosmic time $t_{\rm{end}}$.
In what follows we shall use $e$-folding $\mathcal{N}$ defined as
\begin{align}
~~~\mathcal{N}(t) \equiv \ln \frac{a(t)}{a(t_{\rm{end}})},\label{epsiefol}
\end{align}
and use interchangeably with the cosmic time $t$.
The background energy density is
$\rho_{\mathrm{inf}} = \frac{1}{2} G_{IJ} \dot{\varphi}^I \dot{\varphi}^J + V_E(\varphi^I)$
where $G_{IJ}$ and $V_E$ are evaluated at the background order.

\subsection{Dynamics of the perturbations}
The cosmological perturbations generated by the field fluctuations $\delta\phi^I(x^\mu)$ can account for the observed nearly scale-invariant
adiabatic curvature perturbations. The $\delta\phi^I(x^\mu)$ are in general gauge dependent, however, one can introduce the covariant field fluctuations $\mathcal{Q}^I(x^\mu)$,
which are defined as~\cite{Gong:2011uw,Elliston:2012ab}
\begin{align}
\delta\phi^I &= \mathcal{Q}^I -\frac{1}{2} \Gamma^I_{\ JK} \mathcal{Q}^K \mathcal{Q}^J+\frac{1}{3!} \big(\Gamma^I_{\ MN} \Gamma^N_{\ JK}-\Gamma^I_{\ JK,M}\big)  \mathcal{Q}^K \mathcal{Q}^J  \mathcal{Q}^M+\dots~.
\end{align}
At the linear order, the covariant field fluctuations reduce to the usual field perturbations, i.e., $\mathcal{Q}^I = \delta\phi^I$. One can then express
corresponding gauge-invariant Mukhanov-Sasaki variables as~\cite{Sasaki:1986hm,Mukhanov:1988jd,Mukhanov:1990me}
\begin{align}
Q^I = \mathcal{Q}^I + \frac{\dot{\varphi}^I}{H}\psi = \delta\phi^I+ \frac{\dot{\varphi}^I}{H}\psi.\label{mukh-sasaki}
\end{align}
Note that the variables $Q^I$ are covariant with respect to both space-time and field-space transformations. In the field-space manifold, both $Q^I$ and $\dot{\varphi}^I$ transform as
vectors.
Using Eq.~\eqref{mukh-sasaki} together with Eq.~\eqref{eq:frwmetric} in Eq.~\eqref{eom:scalar}, we get the EoMs for the gauge-invariant perturbations $Q^I$ at linear order
\begin{align}
\mathcal{D}_t^2 Q^\phi &+ 3 H \mathcal{D}_t Q^\phi -\frac{\nabla^2}{a^2} Q^\phi + \mathcal{M}^\phi_{\ \ \phi} Q^\phi = 0,\label{eqpertur:phi}\\
\mathcal{D}_t^2 Q^h &+ 3 H \mathcal{D}_t Q^h -\frac{\nabla^2}{a^2} Q^h + \mathcal{M}^h_{\ \ h} Q^h = 0,\label{eqpertur:h}\\
\mathcal{D}_t^2 Q^{\phi_2} &+ 3 H \mathcal{D}_t Q^{\phi_2} -\frac{\nabla^2}{a^2} Q^{\phi_2} + \mathcal{M}^{\phi_2}_{\ \ \phi_2} Q^{\phi_2}
+ \frac{g_Z}{2} \bigg[\left(\sqrt{\frac{2}{3}}\frac{1}{M_{\rm P}}\right)  \dot{\varphi} h_0 Z_0 -  2 \dot{h}_0 Z_0 + h_0 g_E^{\alpha \nu} (D_\alpha Z_\nu)\bigg]
= 0,\label{eqpertur:phi_2}\\
\mathcal{D}_t^2 Q^{\phi_3} &+ 3 H \mathcal{D}_t Q^{\phi_3} -\frac{\nabla^2}{a^2} Q^{\phi_3} + \mathcal{M}^{\phi_3}_{\ \ \phi_3} Q^{\phi_3}
+ \frac{i e}{2 \sqrt{2} s_W} \bigg[  \left(\sqrt{\frac{2}{3}}\frac{1}{M_{\rm P}}\right)  \dot{\varphi} h_0 \left(W^-_0-W^+_0\right)
- 2 \dot{h}_0  \left(W^-_0-W^+_0\right)\nn\\
&+ g_E^{\alpha \nu} h_0 D_\alpha\left(W^-_\nu-W^+_\nu\right) \bigg]=0,\label{eqpertur:phi_3}\\
\mathcal{D}_t^2 Q^{\phi_4} &+ 3 H \mathcal{D}_t Q^{\phi_4} -\frac{\nabla^2}{a^2} Q^{\phi_4} + \mathcal{M}^{\phi_4}_{\ \ \phi_4} Q^{\phi_4}
+ \frac{i e}{2 \sqrt{2} s_W} \bigg[  \left(\sqrt{\frac{2}{3}}\frac{1}{M_{\rm P}}\right)  \dot{\varphi} h_0 \left(i W^-_0+ i W^+_0\right)
-2 \dot{h}_0  \left(i W^-_0+ i W^+_0\right)\nn\\
&+ g_E^{\alpha \nu} h_0 D_\alpha\left(i W^-_\nu+i W^+_\nu\right) \bigg]=0,\label{eqpertur:phi_4}
\end{align}
with
\begin{align}
\mathcal{M}^{I}_{\ L} = G^{IJ} (\mathcal{D}_L\mathcal{D}_J V_E)- \mathcal{R}^I_{\ JKL} \dot{\varphi}^J \dot{\varphi}^K
- \frac{1}{M_{\rm P}^2 a^3} \mathcal{D}_t \left(\frac{a^3}{H}\dot{\varphi}^I \dot{\varphi}_L\right),\label{eq:massterm}
\end{align}
where $\mathcal{R}^I_{\ JKL}$ is the field-space Riemann tensor. Unless otherwise stated,
all quantities such as $V_E$, $G^{IJ}$, $\Gamma^{I}_{\ JK}$, and $\mathcal{R}^I_{\ JKL}$ are evaluated at the background level.

We first focus on Eq.~\eqref{eqpertur:phi} and Eq.~\eqref{eqpertur:h}.
The two independent field perturbations, $Q^\phi$ and $Q^h$, can be decomposed into adiabatic and isocurvature components. In order to do so, we
first introduce two unit vectors, $\hat{\sigma}^I$ and $\hat{\omega}^I$. The adiabatic unit vector along the
background trajectory in field space is defined as $\hat{\sigma}^I = \frac{\dot{\varphi}^I}{\dot{\sigma}}$ and $\dot{\sigma} = \sqrt{G_{IJ}\dot{\varphi}^I\dot{\varphi}^J}$.
The second unit vector, orthogonal to the background trajectory, is $ \hat{\omega}^I = \frac{\omega^I}{\omega}$, where the turning vector $\omega^I = \mathcal{D}_t \hat{\sigma}^I$
with magnitude $\omega = |\omega^I|= \sqrt{G_{IJ}\omega^I\omega^J}$.

Using these unit vectors, the field perturbations can be projected onto the adiabatic and isocurvature directions as $Q_\sigma = \hat{\sigma}_I Q^I$ and $Q_s = \hat{\omega}_I Q^I$.
The corresponding gauge invariant curvature and isocurvature perturbations can be defined as
\begin{align}
\mathcal{R} = \frac{ H}{\dot{\sigma}} Q_\sigma,~~
\mathcal{S} = \frac{ H}{\dot{\sigma}} Q_s\label{eq:curventropy}.
\end{align}
The dimensionless power spectra associated with the adiabatic and isocurvature perturbations are defined as~\cite{Mukhanov:1990me,Bassett:2005xm,Malik:2008im}
\begin{align}
&\mathcal{P}_{\mathcal{R}}(t;k)= \frac{k^3}{2\pi^2}|\mathcal{R}|^2,
\label{eq:powadia}\\
&\mathcal{P}_{\mathcal{S}}(t;k)= \frac{k^3}{2\pi^2}|\mathcal{S}|^2.
\label{eq:powentrop}
\end{align}
For a given Fourier mode $k$, these quantities can be computed directly from Eqs.~\eqref{eq:powadia} and \eqref{eq:powentrop}. After horizon exit, the curvature perturbation typically approaches a constant value, and hence, the corresponding power spectrum practically freezes. We therefore evaluate $\mathcal{P}_{\mathcal{R}}$ at the end of inflation for the modes of our interest. The isocurvature power spectrum $\mathcal{P}_{\mathcal{S}}(t;k)$, in contrast, can evolve on the superhorizon scales. This evolution arises from the non-vanishing off-diagonal elements of the matrix $M^{I}{}_{J}$, which allow a transfer of power between the adiabatic and isocurvature sectors. Since $\mathcal{P}_{\mathcal{R}}$ is generally much larger than $\mathcal{P}_{\mathcal{S}}$ during inflation, even a mild transfer of power from the adiabatic mode can lead to a noticeable change in the isocurvature power spectrum.
The inflationary observable scalar spectral index is defined as
\begin{align}
n_s = 1 + \frac{d\ln \mathcal{P}_{\mathcal{R}}(k)}
{d\ln k}.
\label{specin1}
\end{align}

The measured values of inflationary parameters are
\begin{align}
 & \log(10^{10}\mathcal{A}_{s*}) = 3.0574 \pm 0.0094~~~(\mbox{CMB-SPA+DESI}~\cite{SPT-3G:2025bzu})\label{obsas}\\
 & n_s^* = 0.9728 \pm 0.0027~~~(\mbox{CMB-SPA+DESI}~\cite{SPT-3G:2025bzu})\label{obsns}\\
 & r_* \lesssim 0.036~\rm{at~95\%~CL}~\cite{BICEP:2021xfz},\label{obs}
\end{align}
where $*$ signifies that  each parameters are measured at the reference scale $k_{\rm ref} = 0.05~\rm{Mpc}^{-1}$. Here $\mathcal{A}_{s*}$ is amplitude of $\mathcal{P}_{\mathcal{R}}$ evaluated  at $k_{\rm ref}$.

\begin{table}[h]
  \centering
  \begin{tabular}{|c|c|c|c|c|c|c|c|c|c|c|c|}
    \hline
    BP  & $\xi_R$ & $\xi_H$ & $\xi_1$  & $\varphi(t_{\text{in}})$ [$M_{\rm P}$] & $h_0(t_{\text{in}})$  [$M_{\rm P}$] & $\mathcal{N}_*$  & $\ln{10^{10} A_{s*}}$ & $n_{s*}$ & $r_*$\\
    \hline
    $a$  & $2.12 \times 10^9$ & 1 & $2\times10^{-9}$  & 5.35 & $10^{-4}$ & -54.16 &  3.054 & 0.9756 & 0.00375 \\
    \hline
    $b$  & $2.32 \times 10^9$ & 10 & $10^{-8}$  & 5.4 & $10^{-5}$ & -56.89  & 3.05948 & 0.9742 & 0.0034 \\
    \hline
  \end{tabular}
  \caption{Benchmark parameter values for $\Phi^2 R^2$ scenario. See text for details.}
  \label{phisqrsq:benchmark}
\end{table}
\begin{table}[h]
  \centering
  \begin{tabular}{|c|c|c|c|c|c|c|c|c|c|c|c|}
    \hline
    BP  & $\xi_R$ & $\xi_H$ & $\xi_2$  & $\varphi(t_{\text{in}})$ [$M_{\rm P}$] & $h_0(t_{\text{in}})$  [$M_{\rm P}$] & $\mathcal{N}_*$ &  $\ln{10^{10} A_{s*}}$ &  $n_{s*}$ & $r_*$\\
    \hline
    1  & $2.18 \times 10^9$ & 1.5 & $5\times10^{-6}$  & 5.37 & $10^{-6}$ & -55 &  3.05537 & 0.9717 & 0.00364 \\
    \hline
    2  & $2.17 \times 10^9$ & 10 & $7\times10^{-6}$  & 5.36 & $10^{-3}$ & -55.03  & 3.0626 & 0.9708 & 0.00362 \\
    \hline
  \end{tabular}
  \caption{Benchmark parameter values for $\Phi^4 R$ scenario.}
  \label{phi4R:benchmark}
\end{table}

\begin{figure}[h]
\begin{center}
\includegraphics[width=.37 \textwidth]{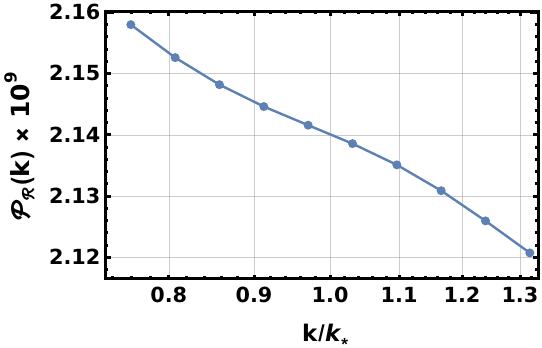}
\includegraphics[width=.37 \textwidth]{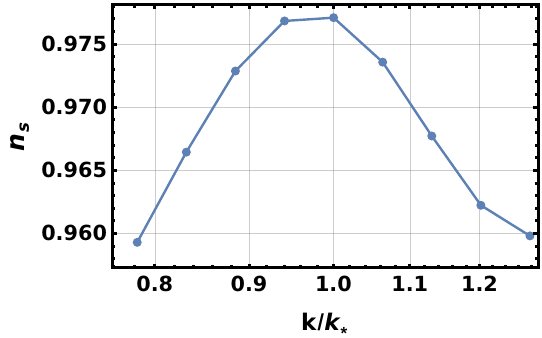}
\includegraphics[width=.37 \textwidth]{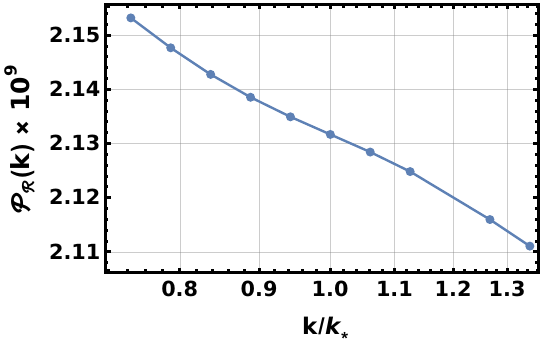}
\includegraphics[width=.37 \textwidth]{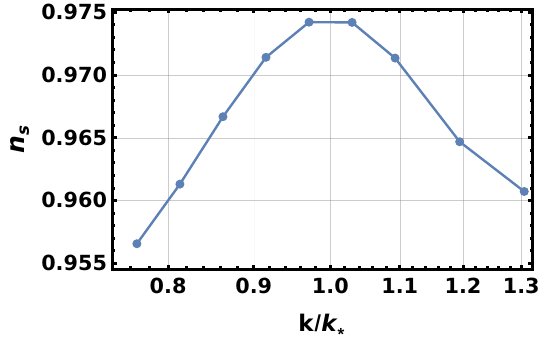}
\end{center}
\caption{The power spectrum $\mathcal{P}_{\mathcal{R}}$ and $n_s$ as a function of $k$ for the $\Phi^2R^2$ scenario with upper and lower panels correspond to BP$a$ and BP$b$ respectively from
Table~\ref{phisqrsq:benchmark}.  See text for details. }
\label{plot:phi2r2}
\end{figure}

\begin{figure}[h]
\begin{center}
\includegraphics[width=.37 \textwidth]{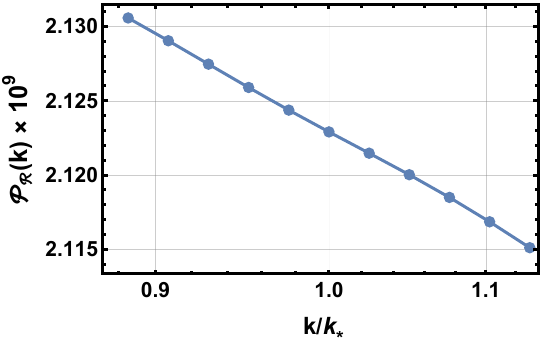}
\includegraphics[width=.37 \textwidth]{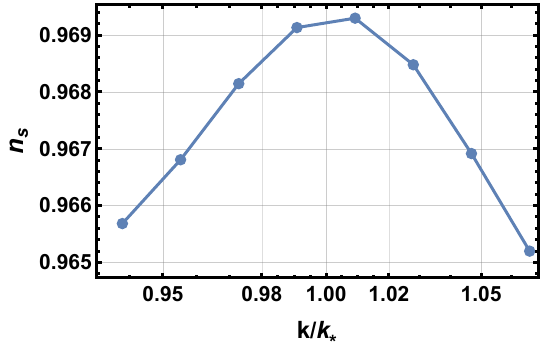}
\includegraphics[width=.37 \textwidth]{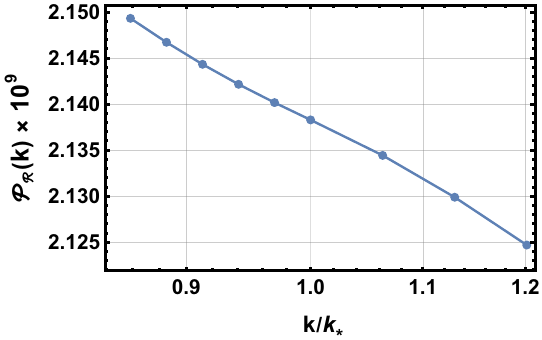}
\includegraphics[width=.37 \textwidth]{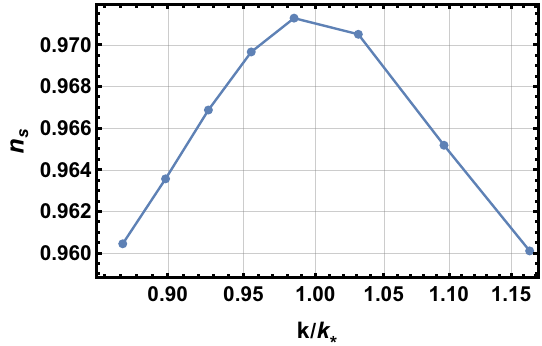}
\end{center}
\caption{Same as Fig~.\ref{plot:phi2r2} but for $\Phi^4R$ scenario.}
\label{plot:phi4R}
\end{figure}

Our primary aim is to study the effects of the nonminimal couplings $\xi_H$, $\xi_1$ and $\xi_2$, and how they help in scale matching via preheating, we consider two benchmark points (BPs) for each of the $\Phi^2R^2$ and $\Phi^4R$ scenarios, summarized in the Table~\ref{phisqrsq:benchmark} and Table~\ref{phi4R:benchmark} respectively. These BPs are found by mildly deviating from the exact single-field-like regime discussed above. In the following analysis, we switch on only one of the dimension-six nonminimal couplings at a time for simplicity as displayed in the respective tables. The dependence of $\mathcal{P}_{\mathcal{R}}$ and $n_s$ for the respective BPs are shown in Figs.~\ref{plot:phi2r2} and \ref{plot:phi4R}, while the corresponding inflationary observables are $n_{s*}$, $\ln{10^{10}(\mathcal{A}_{s*}})$ and $r_*$ are summarized in Table~\ref{phisqrsq:benchmark} and Table~\ref{phi4R:benchmark}. We note that here we have used single-field approximation relation for tensor-to-scalar ratio $r_*\simeq16\epsilon_*$.

\begin{figure}[htbp]
\begin{center}
\includegraphics[width=.37 \textwidth]{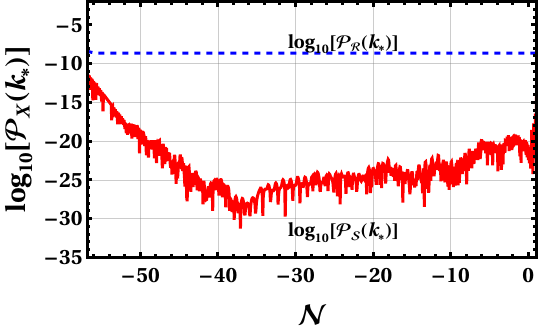}
\includegraphics[width=.36 \textwidth]{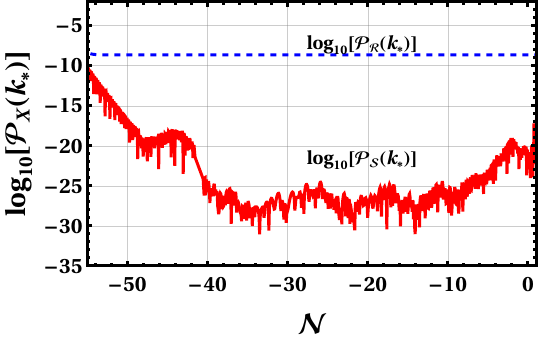}
\end{center}
\caption{The evolution of $\mathcal{P}_{\mathcal{R}}(k_*)$ and $\mathcal{P}_{\mathcal{S}}(k_*)$ for BP$b$ (left) and BP2 (right).}
\label{PRPS}
\end{figure}

\begin{figure}[htbp]
\begin{center}
\includegraphics[width=.42 \textwidth]{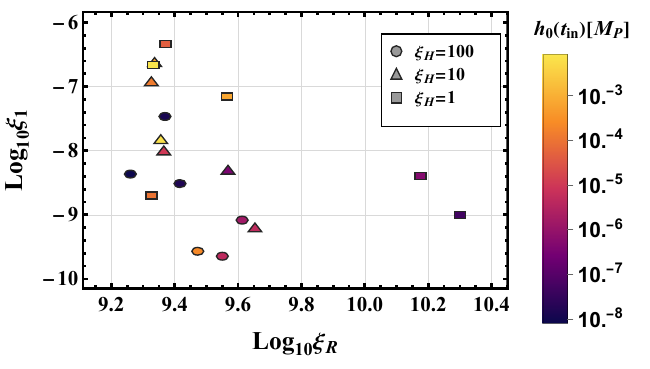}
\includegraphics[width=.42 \textwidth]{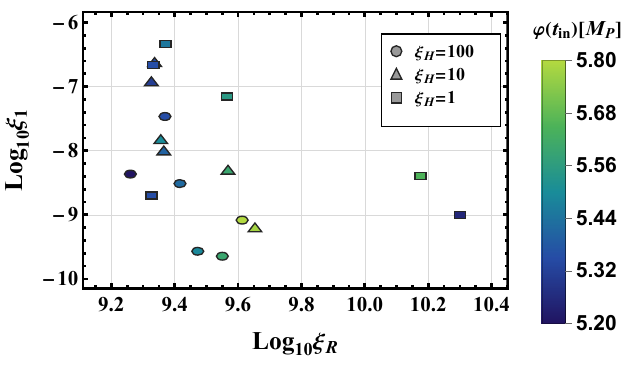}\\
\includegraphics[width=.42 \textwidth]{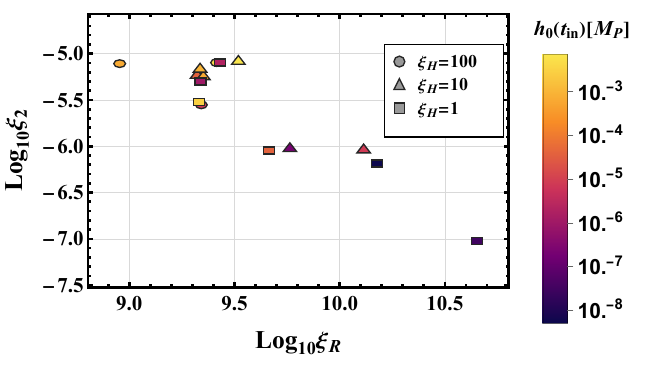}
\includegraphics[width=.42 \textwidth]{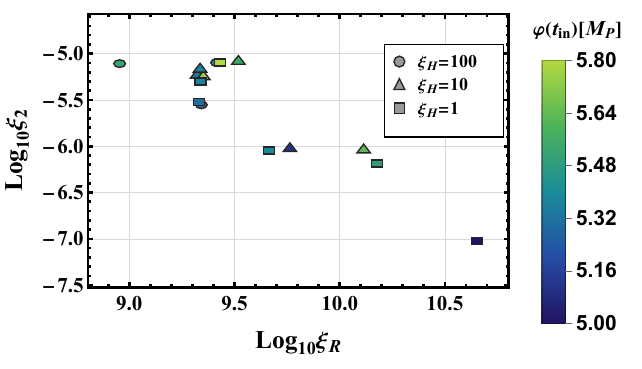}\\
\end{center}
\caption{The scanned parameter space satisfying Eq.~\eqref{obsas}--\eqref{obs} for the $|\Phi|^2R^2$ (upper panel) and $|\Phi|^4R$ (lower panel) scenarios at fixed $\xi_H=1,~10,~100$. In each case, the left panels show the corresponding dependence on $\varphi(t_{\text{in}})$, while the right panels display the same scanned points with dependence on $h_0(t_{\text{in}})$ represented by the color-function bar.}
\label{scanpl}
\end{figure}

The background EoMs of Eq.~\eqref{eq:bkg_inf} are solved with the initial conditions in Table~\ref{phisqrsq:benchmark} and Table~\ref{phi4R:benchmark}. The EoMs of the perturbations in Eq.~\eqref{eqpertur:phi} and Eq.~\eqref{eqpertur:h} on the other hand, are solved in momentum space with Bunch-Davies (BD) initial conditions. The solutions are then plugged into Eqs.~\eqref{eq:powadia} to get $\mathcal{P}_{\mathcal{R}}$ and $n_s$. We find   $n_{s_*}=0.9756$ $(0.9742)$ for BP$a$ (BP$b$) and, 0.9717 and 0.9708 for BP1 and BP2 respectively. It is intriguing that we have accounted for the observed high $n_s$ while $\mathcal{N}_*$ for all BPs remain below 60. Moreover, the values of
$\mathcal{P}_{\mathcal{R}}(k_*)$ are about within $\sim1\sigma$ interval of the CMB-SPA+DESI constraints~\cite{SPT-3G:2025bzu}, whereas and $r_*$ are for all BPs are also consistent with the 95\% CL BICEP/Keck bound~\cite{BICEP:2021xfz}. While our BPs show that indeed one can account for CMB-BAO tension but a full correlation between parameters $\xi_R$, $\xi_H$, $\xi_1$ and $\xi_2$ and $\mathcal{N}_*$ needs a full Bayesian analysis, which is being pursued elsewhere. We refer the reader to Refs.~\cite{Modak:2025bjv,Modak:2025grj} for details. We also find that the $\mathcal{P}_{\mathcal{S}}(k_*)$ is about three orders of magnitude lower than $\mathcal{P}_{\mathcal{R}}(k_*)$ for all BPs and remain much lower during the inflation. This can be easily seen from Fig~\ref{PRPS} where we plot the $\mathcal{P}_{\mathcal{R}}(k_*)$ and $\mathcal{P}_{\mathcal{S}}(k_*)$ as a function of $\mathcal{N}$ for BP$b$ and BP2 which we plot for illustration. However, we still are lacking the matching between $k_*$ and reference scale $k_{\rm ref}$ which is left for the following section section since it depends on the details of the post-inflationary reheating history. Note that the parameter values given in Table~\ref{phisqrsq:benchmark} and Table~\ref{phi4R:benchmark} for both scenarios
are not exact single-field regime rather mild deviation from it. In our $R^2$-like regime (i.e., $\xi_R \gg \xi_H$), the exact single-field solution is found by minimizing the potential $V_E$ with respect to $h_0$ and inserting the minimized value of $h_0$ into $V_E$. Clearly, this is not the case for any of the BPs, as $h_0(t_{\text{in}})$ is nonvanishing in each case.
We shall see shortly that this deviation would help us account for the observed $n_s$ while keeping $\mathcal{N}_* < 60$, by matching the inflationary and CMB reference scales via preheating.
While in Tables~\ref{phisqrsq:benchmark} and~\ref{phi4R:benchmark} we provide four BPs, a finite region of viable parameter space remains for both the $\Phi^2R^2$ and $\Phi^4R$ operators, as illustrated in Fig.~\ref{scanpl}. All scanned parameter points in Fig.~\ref{scanpl} satisfy  the observational constraint in Eq.~\eqref{obsas}--\eqref{obs} and are obtained by mildly deviating from the single-field attractor regime for both the operators as discussed above.

\section{Preheating with Higgs Nonminimal couplings and scale matching}\label{sec:prehea}

\subsection{Preheating dynamics}

In the early part of reheating (i.e. preheating), the rapidly oscillating condensates can produce particles nonperturbatively and modify the thermal history, thereby affect in the matching.
It has been found that Goldstone modes can induce preheating even for $\xi_H\sim 1$ in the $R^2$-like regime in baseline $R^2$-Higgs model if $\xi_H\gtrsim 1$~\cite{Cado:2024von}.
As the BPs in Table~\ref{phisqrsq:benchmark} and Table~\ref{phi4R:benchmark} are also in the similar ballpark range we primarily focus Goldstone preheating while briefly touch on other fields in the following. We begin with first recasting Eqs.~\eqref{eqpertur:phi}--\eqref{eqpertur:phi_4} into a compact form
\begin{align}
&\mathcal{D}_t^2 Q^{I} + 3 H \mathcal{D}_t Q^{I} -\frac{\partial^2}{a^2} Q^{I} + \mathcal{M}^{I}_{\ \ J} Q^{J}+ \mathcal{F}_{(\phi_I)}= 0,\label{eq:perturb}
\end{align}
where $\mathcal{F}_{(\phi)}=\mathcal{F}_{(h)}=0$, $\mathcal{F}_{(\phi_2)}= g_Z [(1/\sqrt{6}M_{\rm P})\,\dot{\phi_0} h_0 Z_0-\dot{h}_0 Z_0+ (h_0/2)(D_\nu Z^\nu)]$, and $g_Z\to i e/(2\sqrt{2}s_W)$ and, $Z_0\to (W^-_0-W^+_0)$ and $(iW^-_0+iW^+_0)$ for $\mathcal{F}_{(\phi_3)}$ and $\mathcal{F}_{(\phi_4)}$, respectively. The $s_W$ denotes the Weinberg angle. As only $\phi$ and $h$ acquire background values, the EoMs of $Q^{\phi_2}$, $Q^{\phi_3}$ and $Q^{\phi_4}$ decouple from $Q^{\phi}$ and $Q^{h}$.

It is clear from Eq.~\eqref{eq:perturb} that the EoMs for the gauge and Goldstone bosons are coupled with each other. Therefore one requires to make a suitable gauge choice to remove some degrees of freedom. The unitary gauge, commonly used in the $R^2$-Higgs model, becomes ill-defined at every zero crossing of the background field $h_0$ after inflation~\cite{Sfakianakis:2018lzf,Cado:2024von}. This is evident from Eq.~\eqref{eq:perturb} and the explicit expression for $\mathcal{F}_{(\phi_2)}$. Likewise, a similar issue appears for $\phi_3$ and $\phi_4$. We therefore use the Coulomb gauge, $\partial_i Z^i=0$ and $\partial_i W^{\pm i}=0$, which does not become ill-defined at zero crossing for the $h_0$~\cite{Sfakianakis:2018lzf,Cado:2024von}. Within this gauge, either the Goldstone fields or the longitudinal gauge bosons may be taken as dynamical variables. We keep the Goldstone bosons as dynamical in the following.

The linear perturbation EoMs, Eq.~\eqref{eq:perturb}, allow us to quantize different fields.
In what follows, we restrict our discussion to $\phi_2$, since the treatment of $\phi_3$ and $\phi_4$ is essentially identical and yields similar results.
The Goldstone boson $Q^{\phi_2}$ is first rescaled as $X^{\phi_2} \equiv a\, Q^{\phi_2}$ and subsequently quantized in momentum space by
\begin{align}
\hat{\widetilde{X}}^{\phi_2} = s_k(\tau) e^{\phi_2}(\tau) \hat{a}(\vb{k}) + s^*_k(\tau) e^{\phi_2}(\tau) \hat{a}^\dagger(-\vb{k}), \label{eq:phi2quant}
\end{align}
where $\hat{\widetilde{X}}^{\phi_2}$ is the Fourier transformed $X^{\phi_2}$ and $\hat{a}$ and $\hat{a}^\dagger$  are the annihilation and creation operators. The $s_k$ is the mode function and,  $e^{\phi_2}$ is the corresponding vielbein, satisfying relation $e^{\phi_2}e^{\phi_2}=G^{\phi_2\phi_2}$ where $\tau$ is the conformal time, related to cosmic time by $\partial_0\to\partial_\tau/a$.

The decoupled mode equation for $\phi_2$ is obtained from Eq.~\eqref{eq:perturb} at the linear perturbation order utilizing quantization relation Eq.~\eqref{eq:phi2quant}~\cite{Sfakianakis:2018lzf,Cado:2024von}
\begin{align}
& s_k'' + \frac{ 2 m_Z^2 \Upsilon}{\mathcal{K}_Z} s_k' + \bigg[ k^2  + a^2 M_{\phi_2}^2+ \frac{ 2 m_Z^2 \Upsilon^2}{\mathcal{K}_Z}\bigg] s_k  = 0,\label{eq:goldmode}
\end{align}
where  $\prime$ denotes the conformal time derivative, $m_Z^2=  (g_Z^2/4)  e^{-\sqrt{\frac{2}{3}}\frac{\varphi}{M_{\rm P}}} h_0^2$,
$M_{\phi_2}^2=m_{\mathrm{eff},(\phi_2)}^2+ m_Z^2$  and,
\begin{align}
&\mathcal{K}_Z =\frac{k^2}{a^2}+ m_Z^2,~~\Upsilon (\tau) = \frac{\varphi'}{\sqrt{6}M_{\rm P}} -  \frac{a'}{a} -   \frac{h_0'}{h_0}.
\end{align}
We remark here that while finding Eq.~\eqref{eq:goldmode} we have rescaled $Z_0 \to Z_0/a$ and utilized the Coulomb gauge condition $\partial_i Z^i=0$.

The effective mass, $m_{\mathrm{eff},(I)}^2$, on the other hand is given as
\begin{align}
m_{\mathrm{eff},(I)}^2(\tau)= \mathcal{M}^I_{~~I} - \frac{1}{6}R_E G^I_{~I}= G^{(I)J} (\mathcal{D}_{(I)}\mathcal{D}_J V_E) - \mathcal{R}^{(I)}_{\ \ JK(I)} \dot{\varphi}^J \dot{\varphi}^K- \frac{1}{M_{\rm P}^2 a^3} \mathcal{D}_t \left(\frac{a^3}{H}\dot{\varphi}^{(I)} \dot{\varphi}_{(I)}\right)-\frac{R_E}{6},\label{eq:meffsq}
\end{align}
where $R_E= -(2 H^2+\dot{H})$.
One reexpress different components of $m_{\mathrm{eff},(I)}^2$ as
\begin{subequations} \begin{eqnarray}
m_{1,(I)}^2 &=& G^{(I)J} (\mathcal{D}_{(I)}\mathcal{D}_J V_E),  \label{def:effective-masses-decomposition-1} \\
m_{2,(I)}^2 &=& - \mathcal{R}^{(I)}_{\ \ JK(I)} \dot{\varphi}^J \dot{\varphi}^K, \label{def:effective-masses-decomposition-2} \\
m_{3,(I)}^2 &=& - \frac{1}{M_{\rm P}^2 a^3} \mathcal{D}_t \left(\frac{a^3}{H}\dot{\varphi}^{(I)} \dot{\varphi}_{(I)}\right), \label{def:effective-masses-decomposition-3} \\
m_{4,(I)}^2&=&-\frac{R_E}{6},  \label{def:effective-masses-decomposition-4}
\end{eqnarray}  \label{def:effective-masses-decomposition}   \end{subequations}
with $(I)$ indices are not summed over, such that
\begin{align}
m_{\mathrm{eff},(I)}^2= \sum_k m_{k,(I)}^2.\label{eq:meffsq1}
\end{align}

The vacuum-subtracted quantum energy density for the Goldstone mode $\phi_2$ is~\cite{Cado:2024von}
\begin{align}
\rho^q_{(\phi_2)} = \frac{1}{a^4}\int\left( \frac{d^3k}{(2\pi)^3} \rho_k^{(\phi_2)} - \frac{k^3}{4\pi^2 \Delta_{(\phi_2)} } dk \right), \label{eq:quantumphi}
\end{align}
where
\begin{align}
\Delta_{(\phi_2)} = \exp{\int_{-\infty}^{\tau}\frac{ 2 m_Z^2 \Upsilon}{\mathcal{K}_Z}\,d\tau'},
\end{align}
and
\begin{align}
\rho_k^{(\phi_2)}\; &= \;   \frac{1}{2} \Biggl\{ \left(1-\frac{m_Z^2}{\mathcal{K}_Z} \right)\left|s_k^\prime\right|^2
 +  \Biggl[k^2+a^2 m_{\mathrm{eff},(\phi_2)}^2 \nn\\
&- \frac{m_Z^2}{\mathcal{K}_Z} \;\Upsilon^2\Biggl] \left|s_k\right|^2
- \frac{m_Z^2}{\mathcal{K}_Z} \;\Upsilon \; (s'_k s^\ast_k +s^{\ast\prime}_k s_k )
\Biggr\}.\label{eq:golphi2}
\end{align}

The quantization and finding the mode equations for Higgs is significantly different since $h$ couples with the inflaton field $\phi$. Therefore, the quantization of these fields needs some
care. The  fluctuations $\hat{\widetilde{X}}^{\phi}(\tau,\vb{k})$ and $\hat{\widetilde{X}}^{h}(\tau,\vb{k})$ are decomposed in momentum space as~\cite{DeCross:2015uza,Cado:2024von}
\begin{align}
\hat{\widetilde{X}}^I(\tau, \vb{k})=  \sum_m \left[ u_m^I(\tau,k) \hat{a}_m(\vb{k}) + u_m^{I*}(\tau,k) \hat{a}^\dagger_m(-\vb{k})   \right],\label{eq:Xmuquantization}
\end{align}
where $m \in \{1,2\}$ associated with  $\{\phi,h\}$. The mode functions $u_m^I(\tau,k)$ are parametrized as
\begin{align}
u_m^I(\tau,k) = t_{(m,I)}(\tau,k) e_m^I(\tau),
\end{align}
where $e_m^I(\tau)$ are vielbeins of the field-space metric and the $t_{(m,I)}(\tau,k)$ are in general complex scalar functions.
The annihilation operators $\hat{a}_m(\vb{k})$ and $\hat{a}^\dagger_m(-\vb{k})$. These are defined as
\begin{align}
 \hat{a}_m(\vb{k})  \ket{0} = 0,\hspace{3 cm} \bra{0} \hat{a}^\dagger_m(\vb{k}) = 0,
\end{align}
and satisfy the usual commutator relationships
\begin{align}
\left[\hat{a}_m(\vb{k}), \hat{a}_n(\vb{q})\right] = \left[\hat{a}^\dagger_m(\vb{k}), \hat{a}^\dagger_n(\vb{q})\right] = 0, \hspace{2 cm}
\left[\hat{a}_m(\vb{k}), \hat{a}^\dagger_n(\vb{q})\right] = (2\pi)^3 \delta_{mn} \delta^{(3)}(\vb{k}-\vb{q}).
\end{align}

We work in practically single-field like regime where the off-diagonal elements $\mathcal{M}^\phi_{~~h}$ and $\mathcal{M}_{~~\phi}^h$ are orders of magnitude smaller that of
$\mathcal{M}^\phi_{~~\phi}$ and $\mathcal{M}^h_{~~h}$ for all four BPs. Hence, the vielbeins also are diagonal and, the rescaled fields in momentum space $\widetilde{X}^\phi$ and $\widetilde{X}^h$ are dependent only on the $v_{1k}(\tau)$ and $y_{2k}(\tau)$, respectively. Hence the mode equations become~\cite{Cado:2024von}
\begin{align}
&v_{1k}'' + \omega^2_{(\phi)}\,v_{1k} \simeq 0,\label{eq:inflatonfluc}\\
&y_{2k}'' + \omega^2_{(h)}\,y_{2k} \simeq 0,\label{eq:higgsfluc}
\end{align}
energy densities of the $\phi$ and $h$ fluctuations per mode are
\begin{subequations} \begin{eqnarray}
\rho_k^{(\phi)} &=& \dfrac{1}{2} G_{\phi\phi}\left(|v'_{1k}|^2 +\omega^2_{(\phi)}   |v_{1k}|^2\right) e_1^\phi e_1^\phi = \dfrac{1}{2}\left(|v'_{1k}|^2 +\omega^2_{(\phi)}   |v_{1k}|^2\right) , \label{enphi_conf-simp}\\
 \rho_k^{(h)} &=& \dfrac{1}{2} G_{hh}\left(|y'_{2k}|^2 +\omega^2_{(h)}  |y_{2k}|^2 \right)e_2^h e_2^h = \dfrac{1}{2}\left(|y'_{2k}|^2 +\omega^2_{(h)}  |y_{2k}|^2 \right), \label{enh_conf-simp}\\
 \rho_k^{\mathrm{int}}& =& \mathcal{O}(h^2)\sim 0.  \label{en-phi+h_conf-simp}
\end{eqnarray}  \end{subequations}
The vacuum subtracted physical quantum energy densities $\rho^q_{(\phi)}$ and $\rho^q_{(h)}$ are found in a similar fashion as in Eq.~\eqref{eq:quantumphi} except with the replacement
$\phi_2\to h$ and $\Delta_{(\phi_2)} \to 1$.

\begin{figure}[h]
\begin{center}
\includegraphics[width=.4 \textwidth]{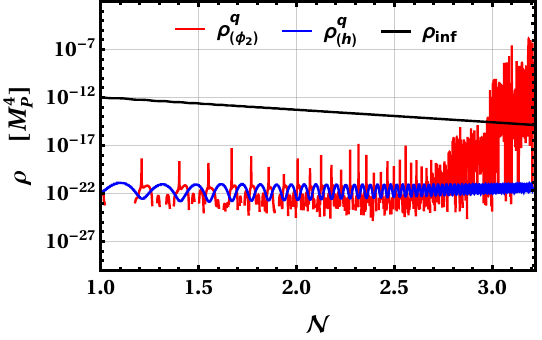}
\includegraphics[width=.4 \textwidth]{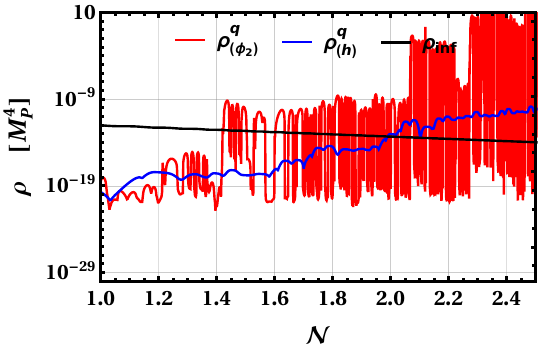}\\
\includegraphics[width=.4 \textwidth]{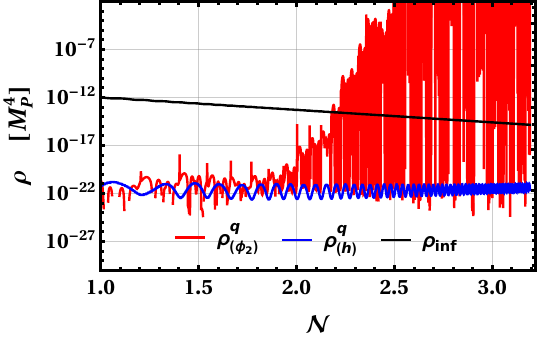}
\includegraphics[width=.4 \textwidth]{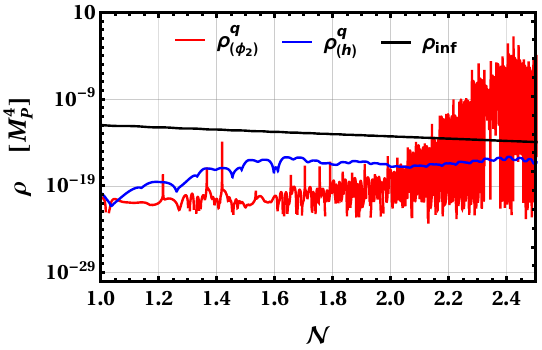}
\end{center}
\caption{The $\rho^q_{(h)}$ (blue) and $\rho^q_{(\phi_2)}$ (red) along with
$\rho_{\rm inf}$ (black) for BP$a$ and BP$b$ in the upper panel and, BP1 and BP2 in the lower panel respectively.}
\label{plot:enerden}
\end{figure}

In Fig.~\ref{plot:enerden}, we show the vacuum-subtracted quantum energy densities of the Higgs field (blue) and the Goldstone boson $\phi_2$ (red), together with the corresponding background energy density (black), for all four BPs. We find that $\rho^q_{(\phi_2)}$ leads to successful preheating in all cases, whereas $\rho^q_{(h)}$ provides successful preheating only for BP$b$. Following Ref.~\cite{Cado:2024von}, preheating is considered complete when the background energy density becomes equal to one of the perturbation energy densities; the corresponding $e$-folding number is denoted by $\mathcal{N}_{\rm pre} = \ln\left(a(t_{\rm pre})/a(t_{\rm end})\right)$. We obtain $\mathcal{N}_{\rm pre}\simeq 3$ for BP$a$, $\mathcal{N}_{\rm pre}\simeq 1.4$ for BP$b$, and $\mathcal{N}_{\rm pre}\simeq 2.2$ for both BP1 and BP2.

\begin{figure}[h]
\begin{center}
\includegraphics[width=.4 \textwidth]{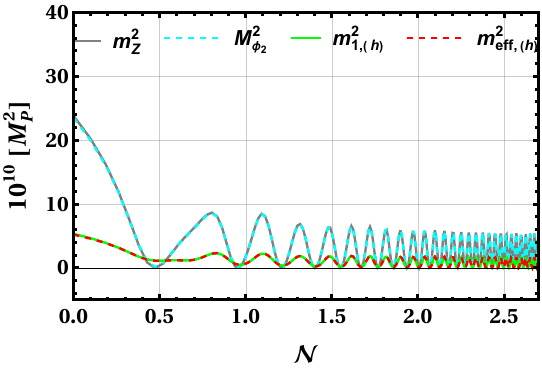}
\includegraphics[width=.4 \textwidth]{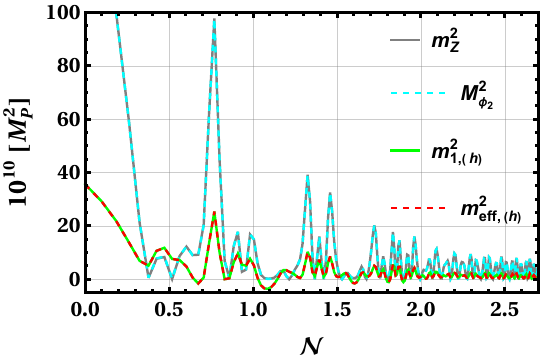}\\
\includegraphics[width=.4\textwidth]{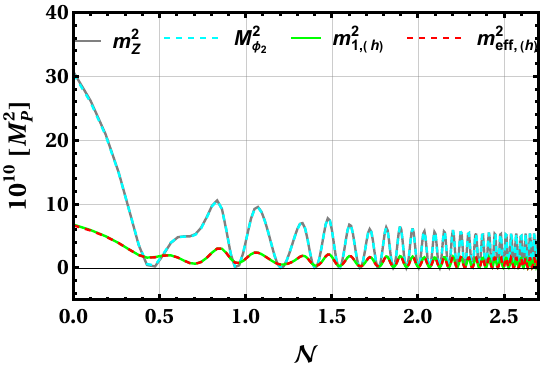}
\includegraphics[width=.4\textwidth]{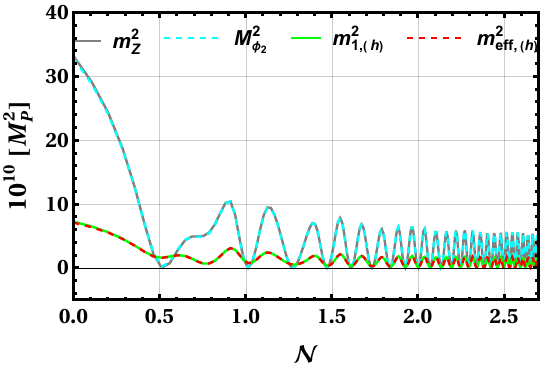}
\end{center}
\caption{The $m_{\mathrm{eff},(h)}^2$ (red dashed) and $M_{\phi_2}^2$ (cyan dashed) for BP$a$ and BP$b$ (upper panel) and, BP1 and BP2 (lower panel) respectively. The respective dominant
contributors to the $m_{\mathrm{eff},(h)}^2$ and $M_{\phi_2}^2$ are $m_{1,(h)}^2$ (green solid) and $m_Z^2$ (gray solid) respectively, shown for all four BPs for comparison. See text for detailed discussion.}
\label{plot:meffplot}
\end{figure}

This behavior is understood from Fig.~\ref{plot:meffplot}, where we plot $M_{\phi_2}^2$ (red dashed) and $m_{\mathrm{eff},(h)}^2$ (cyan dashed) for the different BPs along with their leading contributors. For all BPs, $M_{\phi_2}^2$ remains positive; however, the BP$b$ shows a significantly larger oscillation amplitude amongst them, leading to stronger parametric resonance and as a result leads to faster preheating for the Goldstone mode. In contrast, the oscillation amplitude of $m_{\mathrm{eff},(h)}^2$ is generally mild and cannot generate efficient preheating despite the presence of parametric resonance with the exception of  BP$b$. In case of BP$b$, $m_{\mathrm{eff},(h)}^2$ periodically becomes substantially negative (see the upper right panel of Fig.~\ref{plot:meffplot}), inducing tachyonic instabilities in addition to parametric resonance and therefore provide successful preheating for Higgs.

The findings here are consistent with Ref.~\cite{Modak:2025grj}, which showed that $M_{\phi_2}^2$ is predominantly controlled by $m_Z^2$, while $m_{\mathrm{eff},(h)}^2$ in the single-field regime considered here is mainly determined by $m_{1,(h)}^2$ (see Eq.~\eqref{eq:meffsq1}). We further note that preheating driven by $\rho^q_{(\phi_3)}$ and $\rho^q_{(\phi_4)}$ is qualitatively similar, differing primarily through the replacement of $m_Z^2$ by $m_W^2$. Finally, $\rho^q_{(\phi)}$ stays considerably smaller than the energy densities of the other fields throughout preheating. This is in agreement with Ref.~\cite{Cado:2024von}. If $\xi_1,\xi_2 \to \infty$, the model reduces to the baseline $R^2$-Higgs model. While preheating may still occur for the ballpark $\xi_R$ and $\xi_H$ chosen for the BPs~\cite{Cado:2024von}, these values cannot account for the observed large $n_s$. For $\xi_H<1$, preheating even from the Goldstone bosons becomes suppressed. In such a scenario, nonvanishing $\xi_1$ and $\xi_2$ may still account for the large $n_s$; however, additional fields are required for preheating, and thermalization proceeds via perturbative reheating~\cite{Modak:2025bjv,Ferreira:2025lrd}.

We also remark that the preheating effects, which are byproduct of the parameter space accounting observed high $n_s$, help in scale matching was also identified in $R^3$ modified $R^2$-Higgs inflation~\cite{Modak:2025grj}. It is also worth emphasizing that both the $R^3$ modification and the Higgs-curvature operators $|\Phi|^2R^2$ and $|\Phi|^4R$ can induce indistinguishable inflationary observable and account the observed high $n_s$. The distinction between these scenarios can instead emerge from the subsequent dynamics of the Higgs field $\Phi$. As discussed in Refs.\cite{Kim:2025dyi,Addazi:2025qra,Modak:2025bjv,Park:2025upd,Modak:2025grj}, in the $R^3$ scenario, the Higgs field is not essential for obtaining the modified inflationary predictions, whereas in the Higgs-curvature scenario, dynamics of Higgs plays a direct role in both accounting observed $n_s$ and the scale matching via preheating. Consequently, the post-inflationary evolution of $\Phi$, such as preheating, particle production, and nonlinear dynamics, could in principle distinguish the Higgs-curvature scenario from  the $R^3$ one.

\subsection{The EFT validity and Cut-off scale}

While our analysis is phenomenological, validity of the EFT expansion needs clarification. To determine the validity and corresponding cut-off scale, we expand the action in Eq.~\eqref{eq:actionfinal} around the background $\varphi$ and $h_0$. After canonically rescaling the fluctuations, $ \delta \tilde h = e^{-\frac{\alpha \varphi}{2}} \delta h$ and  $ \delta \tilde \phi_i = e^{\frac{-\alpha \varphi}{2}} \delta \phi_i$ (with $\alpha= \sqrt{\frac{2}{3}}\frac{1}{M_{\rm P}}$) the interaction action takes the form
\begin{align}
 S_{\rm int} \supset &  \int d^4 x \sqrt{-g_E} \bigg(- \frac{1}{2} (\partial \delta \phi)^2 - \frac{1}{2} (\partial \delta \tilde h)^2 - \frac{1}{2}  \sum_{\phi_i} (\partial \delta\tilde\phi_i)^2 + \frac{\alpha}{2} \delta\phi (\partial \delta \tilde h)^2 - \frac{\alpha^2}{4} \delta\phi^2 (\partial \delta \tilde h)^2+\frac{\alpha}{2} \delta\phi (\partial \delta \tilde \phi_i)^2 \nn\\
 &\qquad \qquad \qquad\qquad - \frac{\alpha^2}{4} \sum_{\phi_i} \delta\phi^2 (\partial \delta \tilde \phi_i)^2\bigg),\label{eq:expa}
\end{align}
where $i=2,3,4$ denotes the associated Goldstone bosons. For the scattering $\delta \tilde h \delta \tilde h \to \delta \phi \delta \phi$, from Eq.~\eqref{eq:expa}, we find that the contribution from the four-point kinetic contact interaction terms scale as $A\sim \frac{E^2}{M_{\rm P}^2}$, where $E$ is the typical scattering energy scale. On the other hand, for the $\delta \tilde h \delta \tilde h \to \delta \tilde h \delta \tilde h$ scattering, the contribution from the scalaron exchange diagram scale as $A\sim \left(\frac{E^2}{M_{\rm P}}\right)^2 \frac{1}{E^2}$, as can be seen from Eq.~\eqref{eq:expa}.  These simply translate to unitarity cut-off scale $\sim M_{\rm P}$.

Furthermore, from Eq.~\eqref{eq:actionfinal}, we find the matter-gravity interaction takes the form $\mathcal{L}_{\rm MG}\sim  \frac{1}{M_{\rm P}}\delta{\tilde \zeta_{\mu\nu}} T^{\mu\nu}$, where $T^{\mu\nu}$ is the energy-momentum tensor and $\delta{\tilde \zeta_{\mu\nu}}$ is the canonically normalized metric (graviton) perturbation with $g_{E,\mu\nu}=\bar{g}_{E,\mu\nu}+\frac{1}{M_{\rm P}} \delta{\tilde \zeta_{\mu\nu}}$.
The graviton scattering interactions take typical form of  $\frac{1}{M_{\rm P}} \delta{\tilde \zeta}(\partial  \delta{\tilde \zeta} )^2$ and $\frac{1}{M_{\rm P}^2} (\delta{\tilde \zeta})^2 (\partial \delta{\tilde \zeta} )^2$. Similarly, the corresponding scattering amplitudes also scale as $\sim \frac{E^2}{M_{\rm P}^2}$. Hence, the cut-off scale from the graviton-graviton and matter-gravity scattering is also $\sim M_{\rm P}$. Our findings are consistent with Ref.~\cite{Ema:2016dny} and references therein. Intriguingly, note that, these derivative interactions do not involve the couplings $\xi_1$ and $\xi_2$. For the ranges considered in Table~\ref{phisqrsq:benchmark}, Table~\ref{phi4R:benchmark} and in Fig.~\ref{scanpl}, the $H_*\sim M_\phi\sim\mathcal{O}(10^{-5})M_{\rm P}$ and $V_{E,*}^{1/4}\sim10^{-3}M_{\rm P}$, where $M_\phi^2= \frac{M_{\rm P}^2}{3 \xi_R}$ is the scalaron mass i.e., remain well within the unitarity limit. Further we also find that the magnitude of $m_{\mathrm{eff},(h)}$ and $M_{\phi_2}$ also remain lower than the unitarity cut-off. Likewise, the cut-off scale associated with the interactions of Goldstone bosons also $\sim M_{\rm P}$.

We now turn our attention to the contributions of the process $\delta \tilde h \delta \tilde h \to \delta \tilde h \delta \tilde h$ arising from $V_E$.
Once expanded around the background fields, the scattering receives contributions from the Higgs self-interaction $\lambda (\Phi^\dagger \Phi)^2$ as well as, higher order terms involving $\xi_1$ and $\xi_2$ in $V_E$. This may put additional constraints on the parameters from perturbativity on the effective quartic couplings. Expanding the potential in Eq.~\eqref{eq:potenein} around the background field values and assuming $\frac{h_0^2}{M_{\rm P}^2}\ll 1$  we get the leading contributions for the $\delta \tilde h \delta \tilde h \to \delta \tilde h \delta \tilde h$ scattering as
\begin{align}
V_{(\delta \tilde h)}
&\supset
\frac{1}{4}
\Bigg[
\lambda
+
\frac{1}{\xi_R}
\bigg(
\xi_H^2
-\frac{\left(e^{\alpha \varphi}-1\right)}{2\xi_2}
+\frac{2\left(e^{\alpha \varphi}-1\right)\xi_H}{\xi_1\xi_R}
+\frac{\left(e^{\alpha \varphi}-1\right)^2}{\xi_1^2\xi_R^2}
-15\frac{\xi_H^2}{\xi_1\xi_R}
\frac{h_0^2}{M_{\rm P}^2}
+\frac{15}{2}
\frac{\xi_H}{\xi_2}
\frac{h_0^2}{M_{\rm P}^2}
\bigg)\Bigg]
(\delta \tilde h)^4 ,
\end{align}
which leads to background-dependent perturbativity condition
\begin{align}
\lambda_{\rm eff}(\varphi, h_0) \approx \lambda
+\frac{\xi_H^2}{\xi_R}
-\frac{\left(e^{\alpha \varphi}-1\right)}{2\xi_R\xi_2}
+\frac{2\left(e^{\alpha \varphi}-1\right)\xi_H}{\xi_R^2\xi_1}
+\frac{\left(e^{\alpha \varphi}-1\right)^2}{\xi_R^3\xi_1^2}-15\frac{\xi_H^2}{\xi_1\xi_R^2}
\frac{h_0^2}{M_{\rm P}^2}
+\frac{15}{2}
\frac{\xi_H}{\xi_R\xi_2}
\frac{h_0^2}{M_{\rm P}^2}
\lesssim 4\pi,\label{eq:perut}
\end{align}
In the absence of $\xi_1$ and $\xi_2$ the $\lambda_{\rm eff}$ simply reduces to $\lambda_{\rm UV}=\lambda
+\frac{\xi_H^2}{\xi_R}$ as found in Ref.~\cite{Ema:2016dny} for the baseline $R^2$-Higgs model. Note here that, $\lambda_{\rm eff}$ contain contributions from $h^4$ as well as higher order terms  involving $\xi_1$ and $\xi_2$ in Eq.~\eqref{eq:perut} and are dependent on the background field values $\varphi$ and $h_0$. It is clear from Eq.~\eqref{eq:perut} that during inflation $e^{\alpha \varphi}\gg 1$, however, terms associated with $\xi_1$ and $\xi_2$ are suppressed by factors such as $1/(\xi_R\xi_2)$, $1/(\xi_R^3\xi_1^2)$ and $1/(\xi_1\xi_R^2)$. We find that  $\lambda_{\rm eff}$ remains perturbative for the four BPs. A similar conclusion can be drawn for the Goldstone bosons from scattering $\delta \tilde h \delta \tilde h \to \delta \tilde \phi_i \delta \tilde \phi_i$ and $\delta \tilde \phi_i \delta \tilde \phi_i \to \delta \tilde \phi_i \delta \tilde \phi_i$, for which the background-dependent quartic interactions remain perturbative for the four BPs.

\begin{figure}[h]
\begin{center}
\includegraphics[width=.37 \textwidth]{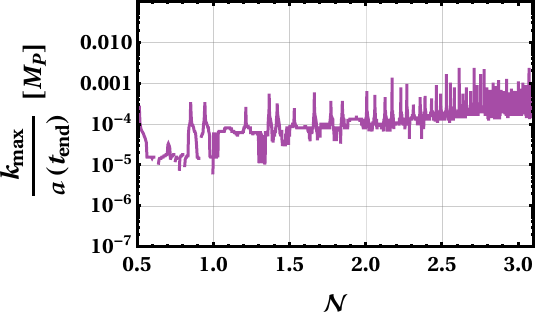}
\includegraphics[width=.37 \textwidth]{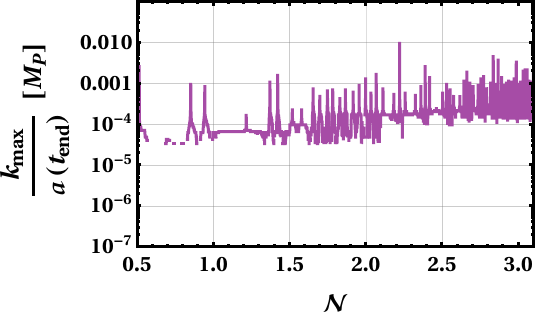}
\end{center}
\caption{The physical momenta of the largest excited  $\phi_2$ Goldstone modes as a function of $\mathcal{N}$ for BP$b$ and BP$2$.}
\label{plot:kmax}
\end{figure}

During preheating, due to its nonperturbative origin, the issue of perturbative unitarity becomes more subtle. Indeed, in case of pure Higgs inflation, longitudinal gauge bosons (or conversely the Goldstone bosons) can be produced violently beyond the unitarity cut-off scale~\cite{DeCross:2015uza,Ema:2016dny,Sfakianakis:2018lzf}. In particular, it has been shown that the if $\xi_H \lesssim 10^3$~\cite{Sfakianakis:2018lzf}, in pure Higgs inflation, produced Goldstone bosons do not violate the unitarity. The situation softens even more for the $R^2$-Higgs inflation~\cite{Cado:2024von}. In Fig.~\ref{plot:kmax} we provide the magnitude of the largest physical momenta $k_{\rm{max}}(t)/a(t_{\rm{end}})$ of the excited modes of the Goldstone bosons $\phi_2$ vs $\mathcal{N}$ for BP$b$ and BP$2$ for illustration~\footnote{Note that here we take $a(t_{\rm{end}})$ in the definition of $k_{\rm{max}}(t)/a(t_{\rm{end}})$ for simplicity. As the scale factor $a(t)$ grows much slowly after the end of inflation our conclusion would remain practically unchanged if we use the definition $k_{\rm{max}}(t)/a(t)$ instead of $k_{\rm{max}}(t)/a(t_{\rm{end}})$.}. It is clear that, before the completion of preheating, at each time step, the $k_{\rm{max}}/a(t_{\rm{end}})\ll M_{\rm P}$. This illustrates that the produced modes do not violate perturbative unitarity.  Note, however, this conclusion is based on linear order approximation only applicable for modes produced before completion of preheating.

\subsection{Scale Matching}
We now proceed towards matching. Assuming that  the thermalization is immediately completed after preheating, we estimate preheating temperature $T_{\rm pre}$ via
\begin{align}
\rho_{\mathrm{inf}}\bigr|_{\mathcal{N}=\mathcal{N}_{\rm pre}}\equiv\rho_{\rm pre}=\frac{g_{\rm pre} \pi^2}{30} \; T_{\rm pre}^4,\label{eq:prehettemp}
\end{align}
where $g_{\rm{pre}}=106.75$ is the number of relativistic degrees of freedom at the completion of preheating and $T_{\rm pre}$ is the preheating temperature.
We find that $T_{\rm pre} \approx 2.3 \times 10^{14}~(7.5 \times 10^{14})$ GeV for BP$a$ (BP$b$) and, $\approx 4.2~(4.4\times 10^{14})$ GeV for BP1 (BP2).
The next step is to relate the reference mode $k_*$, expressed in units of $M_{\rm P}$, to the CMB pivot scale $k_{\rm ref}/a_0 = 0.05~{\rm Mpc}^{-1}$. The reference mode $k_{\rm ref}$ is related to $k_*$ through
\begin{align}
k_{\rm ref} =  k_* = a(t_*) H(t_*)= \frac{a(t_*)}{a(t_{\rm end})} \frac{a(t_{\rm end})}{a(t_{\rm pre})} \frac{a(t_{\rm pre})}{a_0} a_0 H(t_*), \label{eq:kefrel}
\end{align}
where $a(t_{\rm pre})$ denotes the scale factor at the end of the preheating. We assume that the radiation dominated epoch begins immediately after preheating.
We can now reexpress Eq.~\eqref{eq:kefrel} as~\cite{He:2020ivk}
\begin{align}
\mathcal{N}_* = -\ln\Biggl[\frac{H_*}{k_{\rm ref}/a_0} \frac{T_0}{T_{\rm pre}} \frac{g_0^{1/3}}{g_{\rm pre}^{1/3}}\Biggr] + \mathcal{N}_{\rm pre}, \label{eq:nstar}
\end{align}
where  $T_0=2.7 K$  is the temperature today and $g_0= 43/11$ is the relativistic degrees of freedom today. It should be noted that while deriving Eq.~\eqref{eq:nstar}, we have assumed that thermalization completes within one Hubble time~\cite{He:2020ivk}.

Converting all dimensionful quantities in $M_{\rm P}$, the right hand side of Eq.~\eqref{eq:nstar} yields -54.94 and -55.31 for the BP$a$ and BP$b$ respectively, whereas, -55.22 and -55.07 for BP1 and BP2. This should be compared with the left hand side i.e. of Eq.~\eqref{eq:nstar} summarized in Table~\ref{phisqrsq:benchmark} and Table~\ref{phi4R:benchmark}. This shows that the two sides agree within $\sim1$--1.5 $e$-folding for the BPs, but, there is a subtlety. Indeed, one can readjust the input parameters $\xi_R$, $\xi_H$, $\xi_1$, and $\xi_2$, as well as the initial field values, to achieve an exact matching of both sides of Eq.~\eqref{eq:nstar} for the BPs.  We refrain from doing so here, primarily because our linear order analysis does not include condensate and perturbation decays, backreaction or rescattering effects of the produced particles. Furthermore, we have also checked that for the scanned points in Fig.~\ref{scanpl}, preheating from Goldstone and Higgs also help in scale matching, albeit mild uncertainties as mentioned above. We also remark that there may exist regions of parameter space away from the single-field-like regime shown in Fig.~\ref{scanpl}; however, the preheating dynamics in these regions do not help in matching the relevant scale.

The above mentioned effects will provide some uncertainties in the  thermalization  process and the equation of state of the system may deviate from 1/3 and hence the matching relation of Eq.~\eqref{eq:nstar}  may acquire additional correction (see e.g. Ref.~\cite{Cheong:2021kyc} for similar discussion).
Consequently, this uncertainty may affect the $\mathcal{N}_*$ and hence the predicted $n_s$. The uncertainty can be estimated simply by scaling the energy density at the end of preheating to the energy density at the thermalization via
\begin{align}
 \rho_{\rm th} = \rho_{\rm pre} \left(\frac{a_{\rm th}}{a_{\rm pre}}\right)^{-3\left(1+ \bar{\omega}\right)} = \rho_{\rm pre} \ e^{-3\left(1+ \bar{\omega}\right)\mathcal{N}_{\rm th}},
\end{align}
where $\bar{\omega}$ is the average equation of state, $\mathcal{N}_{\rm th}= \left(\frac{a_{\rm th}}{a_{\rm pre}}\right) $ and $a_{\rm th}$
is the scale factor at the thermalization. For simplicity, by defining $\rho_{\rm pre}=\frac{g_{\rm pre} \pi^2}{30} \; T_{\rm pre}^4$ and $\rho_{\rm th}=\frac{g_{\rm th} \pi^2}{30} \; T_{\rm th}^4$ and, assuming $g_{\rm th}=g_{\rm pre}$ one can find $\Delta\mathcal{N}_{\rm th} \simeq \frac{1- 3\bar{\omega}}{4}\mathcal{N}_{\rm th}$.

The $\Delta\mathcal{N}_{\rm th}$ would lead to some uncertainty in $n_s$. Assuming $\Delta\mathcal{N}_{\rm th}\ll \mathcal{N}_*$, and as all BPs practically remain in the single-field like regime, this uncertainty amounts a shift
\begin{align}
\Delta n_s
\simeq
\frac{1-3\bar{\omega}}{2\mathcal{N}_*^2}
\mathcal{N}_{\rm th}.
\end{align}
We notice that for instantaneous thermalization, i.e., $\mathcal{N}_{\rm th}=0$ the $\Delta\mathcal{N}_{\rm th} \to 0$, and hence $\Delta n_s \to 0$.
In Table~\ref{tab:dns_thermalization}, we provide the magnitude of the associated uncertainties $|\Delta n_s|$ for different values of $\bar{\omega}$ and $\mathcal{N}_{\rm th}$ for BP$a$ and BP1 for illustration. Similar $\Delta n_s$ is found for the other two BPs.
\begin{table}[htbp]
\centering

\begin{tabular}{|c|c|c|c|c|}
\hline
$\bar{\omega}$
& $\mathcal{N}_{\rm th}=1$
& $\mathcal{N}_{\rm th}=3$
& $\mathcal{N}_{\rm th}=5$
& $\mathcal{N}_{\rm th}=10$ \\
\hline
$-\frac{1}{3}$ & 0.00033 (0.00033)  & 0.00099 (0.00098) & 0.00166 (0.00164) & 0.00331 (0.00328) \\
$0$            & 0.00017 (0.00016)  & 0.00049 (0.00049) & 0.00083 (0.00082) & 0.00166 (0.00164) \\
$\frac{1}{6}$  & 0.00008 (0.00008)  & 0.00025 (0.00025) & 0.00041 (0.00041) & 0.00083(0.00082) \\
$\frac{1}{3}$  & 0  & 0 & 0 & 0 \\
\hline
\end{tabular}
\caption{The shift in the scalar spectral index $|\Delta n_s|$ for BP$a$ (BP1)
different values of the effective equation-of-state parameter
$\bar{\omega}$ and the duration of the thermalization phase
$\mathcal{N}_{\rm th}$.}
\label{tab:dns_thermalization}
\end{table}
It is clear that for larger $\mathcal{N}_{\rm th}$, $\Delta n_s$ increase, however, stay well within the $2\sigma$ error bar of the observed $n_s$ as given in Eq.~\eqref{obsns} even for $\mathcal{N}_{\rm th}=10$.

\section{Discussion and Summary}\label{disc}

The recently observed higher value of $n_s$ by ACT and SPT, together with BAO measurements from DESI, disfavors both pure $R^2$ inflation and the single-field regime of $R^2$-Higgs inflation. We have shown that the inclusion of the dimension-six operators $\left|{\Phi}\right|^2R_J^2$ and $\left|{\Phi}\right|^4R_J$ within the $R^2$-Higgs model can successfully accommodate the observed cosmological data. The associated couplings $\xi_1$ and $\xi_2$, along with $\xi_R$ and $\xi_H$, can lower the required number of $e$-folding to $\mathcal{N}_*<60$, while also reproducing the observed value of $n_s$. Remarkably, the same region of parameter space naturally gives rise to efficient preheating, which helps establish a consistent matching between the reference CMB scale and the inflationary scale. The preheating dynamics are dominated by Goldstone bosons (equivalently the longitudinal gauge boson modes), emphasizing their crucial role in the post-inflationary evolution of the Universe.

We remark that these preheating effects have not been discussed in $R^2$-Higgs inflation in the context of the ACT and SPT results. The primary reason is the treatment of the Standard Model gauge sector. Nearly all studies of $R^2$-Higgs inflation following the ACT and SPT data adopted the unitary gauge, which removes the longitudinal gauge degrees of freedom from the dynamics. In contrast, here we employ the Coulomb gauge, which is the appropriate gauge choice for studying preheating~\cite{Sfakianakis:2018lzf,Cado:2024von}. In the context of the ACT and SPT results, Ref.~\cite{Modak:2025grj} was the first to identify that the Goldstone bosons can efficiently preheat the Universe while simultaneously accounting for the observed $n_s$.

In general $f(R)$ theories, it has also been suggested that an $R^3$ term may account for the observed $n_s$~\cite{Kim:2025dyi,Addazi:2025qra,Modak:2025bjv,Park:2025upd,Modak:2025grj}. Furthermore, other nonminimal curvature couplings involving $R_{\mu\nu}R^{\mu\nu}$, $R_{\mu\nu\rho\sigma}R^{\mu\nu\rho\sigma}$, as well as the Gauss--Bonnet combination $R^{2}-4R_{\mu\nu}R^{\mu\nu}+R_{\mu\nu\rho\sigma}R^{\mu\nu\rho\sigma}$, and their dimension-six analogues, are also possible and may lead to modifications of the inflationary predictions~\cite{Satoh:2007gn,Weinberg:2008hq,Guo:2009uk,Odintsov:2018zhw,Nojiri:2019dwl,Kawai:2021edk,Koh:2023zgn,Nojiri:2026hij,Odintsov:2026cxz,Odintsov:2026dss}. In addition, one may have dimension-six $\frac{\gamma}{M_{\rm P}^2}|\Phi|^6$ operator, which may also impact on the inflationary observables as discussed in Ref.~\cite{Lee:2023wdm}, this operator can have a significant impact on the inflationary observables if $\gamma \simeq \mathcal{O}(10^{-3}$)--($10^{-6})$ for $\lambda \simeq 10^{-2}$--$10^{-4}$. We remark that such operator could potentially account for the observed high $n_s$ which may add further degeneracies to $(\xi_R,\xi_H,\xi_1,\xi_2)$ along with dimension-six $R^3$ term.

While the observed CMB-BAO tension may be hinting at the presence of higher dimensional operators, some caution is needed. It has been discussed that the correlations among the datasets need to be better understood within the baseline $\Lambda$CDM model and that the tension could simply be the result of unaccounted systematic effects~\cite{Ferreira:2025lrd}. Nevertheless, the tension is intriguing. If the tension persists and is eventually established as evidence for physics beyond the Standard Model, it could provide a new window into quantum effects of gravity appearing from higher-curvature operators, at least at the linear level.

\subsection*{Acknowledgments}
AS thanks Bikash Kumar Acharya for discussion.

\renewcommand{\emph}{}
\bibliography{references}

\end{document}